\documentclass{PSSOH} 
\usepackage{newtxtext,newtxmath}
\usepackage{multirow}
\usepackage{tabularx}
\usepackage{array}
\usepackage{soul}
\usepackage{lscape}
\usepackage{longtable}
\newcolumntype{L}[1]{>{\raggedright\let\newline\\\arraybackslash\hspace{0pt}}m{#1}}

\begin{document}
\addtocounter{page}{89}

\centerline{PSSOH 2021, DOI: \href{https://doi.org/10.5281/zenodo.5524414}{https://doi.org/10.5281/zenodo.5524414}}
\hfill \break

\title{TOWARDS FAIR PRINCIPLES FOR OPEN HARDWARE} 

\author{
    Nadica Miljkovi\'{c}$^1$, 
    Ana Trisovic$^2$, 
    Limor Peer$^3$
    }

\date{
1: School of Electrical Engineering, University of Belgrade, Belgrade, Serbia\\
2: Institute for Quantitative Social Science, Harvard University, Cambridge, MA, USA\\
3: Institution for Social and Policy Studies, Yale University, New Haven, CT 06520, USA\\
E-mail addresses: \href{mailto:nadica.miljkovic@etf.bg.ac.rs}{nadica.miljkovic@etf.bg.ac.rs}, \href{mailto:anatrisovic@g.harvard.edu}{anatrisovic@g.harvard.edu}, \href{mailto:limor.peer@yale.edu}{limor.peer@yale.edu}
}

{\let\newpage\relax\maketitle}

\maketitle

\begin{abstract}
The lack of scientific openness is identified as one of the key challenges of computational reproducibility. In addition to Open Data, Free and Open-source Software (FOSS) and Open Hardware (OH) can address this challenge by introducing open policies, standards, and recommendations. However, while both FOSS and OH are free to use, study, modify, and redistribute, there are significant differences in sharing and reusing these artifacts. FOSS is increasingly supported with software repositories, but support for OH is lacking, potentially due to the complexity of its digital format and licensing. This paper proposes leveraging FAIR principles to make OH findable, accessible, interoperable, and reusable. We define what FAIR means for OH, how it differs from FOSS, and present examples of unique demands. Also, we evaluate dissemination platforms currently used for OH and provide recommendations.

\end{abstract}

\section{Introduction}

Open science emerged as a movement to make scientific research available to broad audiences, from professionals to the general public~\cite{dryden_upon_2017, hocquet_epistemic_2021, wheeler2011free}. In particular, scientific publications, data, physical samples, and software should be made transparent and accessible whenever possible~\cite{fortunato_case_2021, marwick2017computational, martinez2018reproducibility}. 
The movement, helped by community-driven efforts such as the Turing Way~\cite{community_turing_2019} and Global Open Science Hardware (GOSH)~\cite{gosh}, includes practices like open access to published research, releasing software as Free and Open-Source (FOSS), and experimental instruments as Open-Source Hardware or Open Hardware (OH).\footnote{In this paper, we use a common OH abbreviation, however FOSH (standing for Free and Open-Source Hardware) is also occasionally used.} These open practices aim to facilitate scientific verification, reuse, and collaboration and to inspire trustworthiness in science.

Software and hardware have been an integral part of scientific research and are increasingly recognized in academic journals and conferences that often encourage their dissemination upon publication. FOSS is, by definition, software that "respects users' freedom and community," which means that it adheres to four essential freedoms: to run the program, to study how the program works, to redistribute copies, and to distribute the modified copies (commercially or non-commercially)~\cite{freesw, stallman2002free}. OH is defined as a "physical artifact, either electrical or mechanical, whose design information is available to, and usable by, the public in a way that allows anyone to make, modify, distribute, and use" it~\cite{tapr1, ohsource}. It represents a set of design and legal principles and can refer to a wide range of objects such as computers, scientific instruments, 3D printed furniture, physical constructions, and robots. In practice, OH is commonly captured as digital schematics and drawings with design instructions and a license that allows a reuser to construct and put it to use.\footnote{In this paper, we examine OH used for research, which in addition to the design component typically includes  software.}

The OH concept was modeled on free software, which led to many issues as software and hardware are fundamentally different and comply with different standards and design principles~\cite{bonvoisin2017source}. Reported challenges of OH reuse include high costs in its construction, imprecise documentation, and complex licensing~\cite{pearce_return_2016, rubow2008open, ackerman2008toward, ohsource}. In particular, the cost of reusing OH can be high as it implies building custom-made physical artifacts at non-negligible expenses, while in contrast, the cost of reusing FOSS is often marginal.\footnote{OH can be highly profitable~\cite{pearce_return_2016, HEIKKINEN2020119986}. For example, opening Arduino OH and FOSS created significant revenue of \$56.8~M per year (https://growjo.com/company/Arduino).} Even slight imprecision in OH description, such as resistor power in schematics, can lead to an unintended final product and a failed investment. Also, OH licenses are complex and have the added risk of patent infringement because OH specifications encompass various artifacts, including design schemes and simulation codes~\cite{rubow2008open}. Because of these issues, reusing or reproducing a study based on OH can be particularly challenging. Reproducibility refers to "obtaining consistent results using the same input data, computational steps, methods, and code, and conditions of analysis" ~\cite{stodden_enabling_2018, engineering2019reproducibility, fortunato_case_2021, peer_challenges_2021}. The need for reproducibility led to a surge of research and development, making scientific work more streamlined, but the challenges of OH reuse still remained.

In this paper, we propose an application of FAIR principles~\cite{wilkinson_fair_2016} that could help alleviate the complexities in OH dissemination and reuse. We modify FAIR principles to incorporate the needs of OH users. Moreover, we examine current dissemination platforms and evaluate their effectiveness for OH. Our recommendations support open-source values and practical implementations that make OH more discoverable, reusable, and transparent. They should be of interest to OH users, scientists, and repository managers.

\begin{figure}
    \centering
    \includegraphics[width=0.7\linewidth]{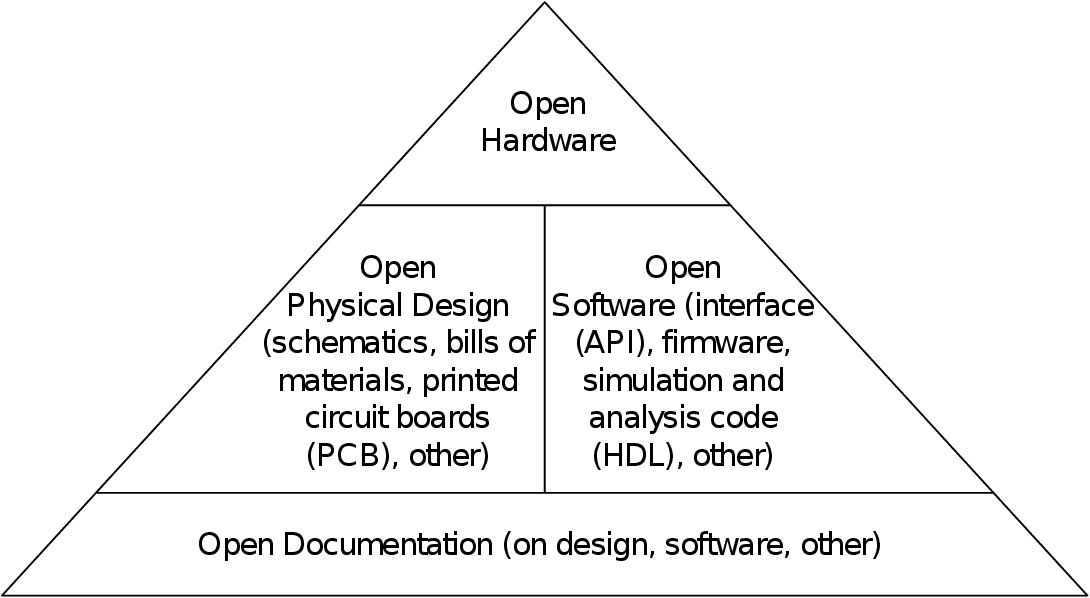}
    \caption{Components of OH (inspired by OSHWA~\cite{rubow2008open, ackerman2008toward}).}
    \label{fig:oh-schema}
\end{figure}

\section{Open Hardware: background, use cases, and challenges}

We can distinguish three components of OH digital form: physical design, software, and documentation (Fig.~\ref{fig:oh-schema}). The physical design includes mechanical drawings, connectivity diagrams (schematics), bills of materials, printed circuit boards (PCB), layout data, and more. The software component may include HDL (hardware description language) source code, which is used to simulate and validate the design's intended functionality while allowing architectural exploration and comparing variations of a base design. The software component may also include software firmware (drivers) that operate the hardware, an open interface to the hardware, and an open implementation covering a set of tools to create and test a design. Finally, the documentation component offers information on assembling and running the OH. Understanding the structure of OH helps us grasp the complexity of its capture, dissemination, and reuse.

The use of OH in science has been recognized internationally by researchers and institutions. The nonprofit Open Source Hardware Association (OSHWA) fosters technological knowledge of hardware, promotes its development, and maintains certification. The increased popularity of OH is demonstrated by the significant rise of certified projects over a single year (64\%, from 977 on August 24, 2020 to 1601 on July 19, 2021)~\cite{pejovic_predrag_2021_4748368}. It is estimated that a national OH policy in Finland (with a funding mechanism that supports OH development) would provide 90\% in savings compared to the cost of proprietary hardware~\cite{HEIKKINEN2020119986}, which is in line with the EU findings~\cite{eurep}. The first official government document with a strategy for FOSS in research for 2021-2024 was released in July 2021 by the French Ministry of Higher Education, Research, and Innovation~\cite{frnational}. Another example is the European Organization for Nuclear Research (CERN) introducing a policy to publish its experimental hardware in open access journals and invest in OH initiatives~\cite{cernoh}. 

\begin{figure}[ht]
    \centering
    \includegraphics[width=0.85\linewidth]{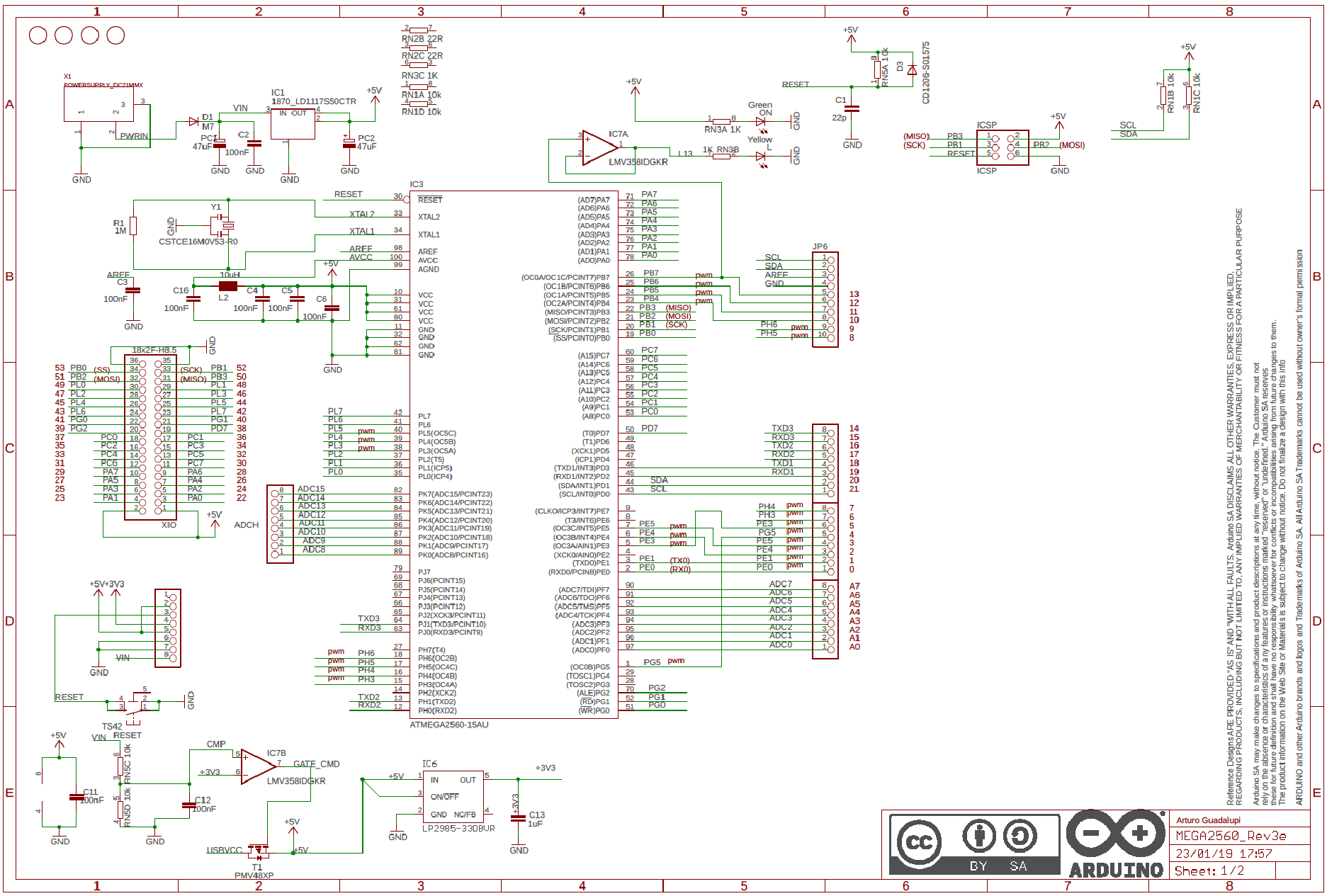}
    \caption{Arduino MEGA 2560 Schematics licensed under CC BY SA license~\cite{mega25}}
    \label{fig:arduino}
\end{figure}

\begin{figure}
    \centering
    \includegraphics[width=0.65\linewidth]{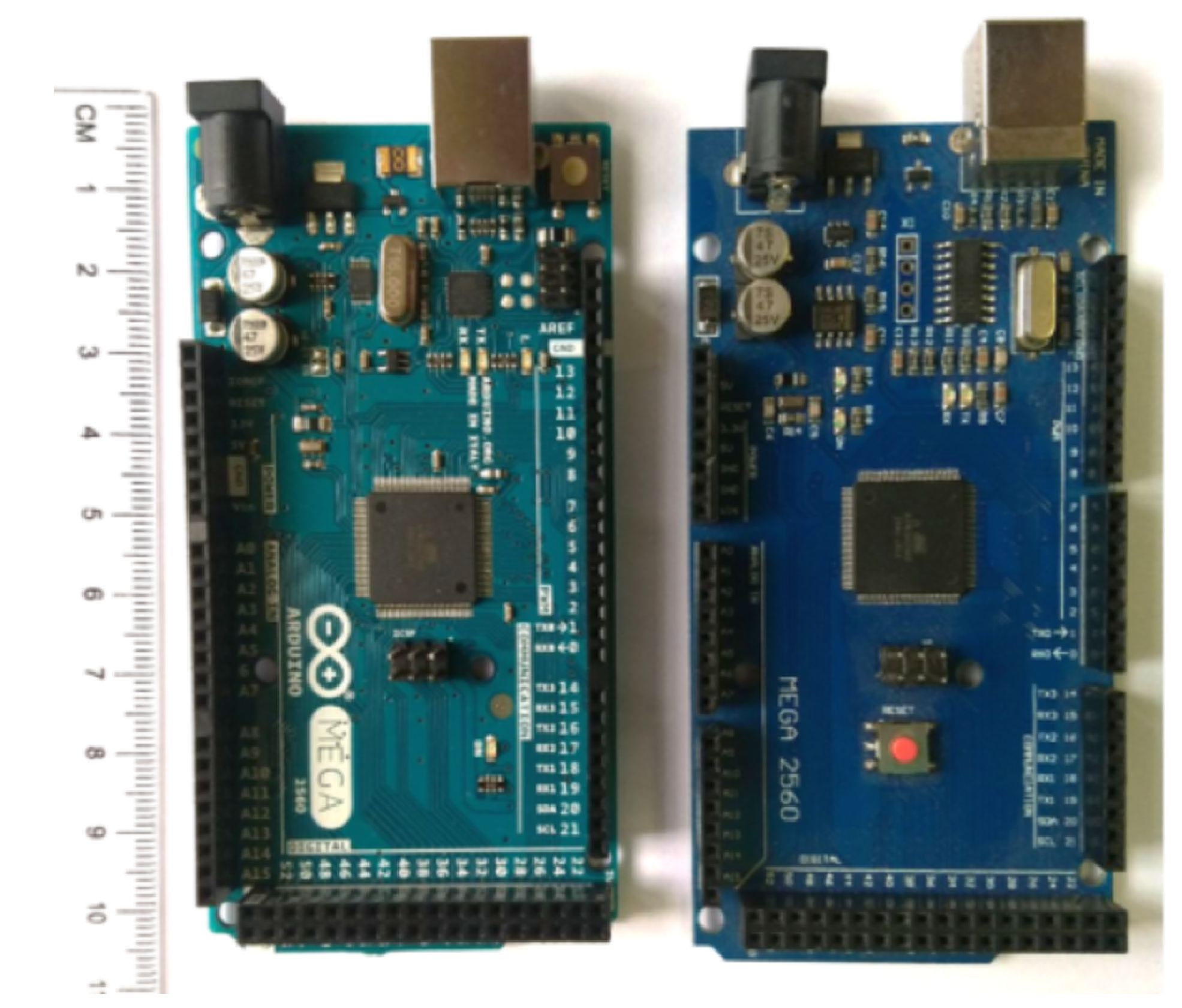}
    \caption{Comparison of the original Arduino MEGA 2560 board produced in Italy on the left, and a derived MEGA 2560 board produced in China based on the open design (see schematics at Fig.~\ref{fig:arduino}). Photo Credit: N.M.}
    \label{fig:photo}
\end{figure}

An example of a highly successful and widely used OH design is the Arduino MEGA 2560 microcontroller board (screenshot of the design file shown in Fig.~\ref{fig:arduino}). With the open design file and its permissive license, a replica (Fig.~\ref{fig:photo} right) of the original printed circuit board (Fig.~\ref{fig:photo} left) can be produced and legally sold around the world~\cite{mega25}. However, the copies cannot contain the Arduino name and logo as the trademark is protected. The trademark can be used if a company becomes an official manufacturer, such as Smart Projects in Italy, SparkFun in the USA, and Dog Hunter in Taiwan/China~\cite{requiredbrain}. These boards are easily programmable and able to receive input (e.g., read input from a light sensor) and generate output according to a custom algorithm (e.g., if a room is dark, turn on the light) at a low price (e.g., MEGA 256 board costs at most 35 EUR~\cite{arduino1}). In addition to simple applications, Arduino boards have been extensively used in scientific research~\cite{HEIKKINEN2020119986, pearce2012building}, including as part of other OH designs~\cite{KUMBOL2018e00047, MNATI2021e00183, bioamp}.

To illustrate the challenges of OH dissemination and reuse in scientific research, we examine three published studies that used OH: Actifield device for measuring movements of laboratory rodents~\cite{KUMBOL2018e00047}, a non-contact thermometer to prevent the spread of the contagious diseases~\cite{MNATI2021e00183}, and an electronic board BioAmp EXG Pill for measuring signals in the human body such as electrical heart potentials or electrocardiogram (ECG)~\cite{bioamp}. 

Three dedicated licenses were applied for the BioAmp EXG Pill design, software, and documentation, as mandated for OSHWA certification (Tab.~\ref{tab:use-cases}). Projects disseminated through HardwareX journal are assigned the same license for both hardware and software components: for the thermometer, a CC BY-SA license was applied, and for the actifield device, a GNU GPL license was used. The CC license is mainly intended for work regulated by copyright law (e.g., for books, music, photographs, articles) and therefore is less suitable for software and OH. On the other hand, the GNU GPL license is widely used for FOSS, but it is ineligible for OH. The documentation was licensed under CC BY in all use cases, but some differences emerged. For example, the actifield device and the thermometer HardwareX articles provide more details for further OH reuse due to the article template requiring information such as build and operation instructions, validation and characterization, and the bill of materials~\cite{todd_duncombe_2021_5078227}. Another advantage of dissemination through specialized journals such as HardwareX is the application of persistent identifiers. Although OSHWA also provides unique identifiers, information on their persistence is not available.

\begin{table}[ht]
    \centering
    \footnotesize
    \begin{tabular}{llccc}
        \toprule\toprule
        \multicolumn{2}{l}{\textbf{Use case}} & \textbf{Actifield device}~\cite{KUMBOL2018e00047} & \textbf{Thermometer}~\cite{MNATI2021e00183} & \textbf{BioAmp EXG Pill}~\cite{bioamp} \\ \midrule
        \multicolumn{2}{l}{\textbf{Main dissemination channel}} & \multicolumn{2}{c}{HardwareX journal} & OSHWA \\ \midrule
        \multirow{3}{*}{\textbf{Licenses}} & \textbf{Physical design} & \multirow{2}{*}{GNU GPL} & 
        \multirow{2}{*}{CC BY-SA} & CERN-OH\\ 
        \cmidrule{5-5}
         & \textbf{Software} & & & MIT \\ \cmidrule{3-5} 
         & \textbf{Documentation} & \multicolumn{3}{c}{CC BY} \\ \midrule
        \multicolumn{2}{l}{\textbf{Design files repository}} & \multicolumn{2}{l}{Open Science Framework (OSF)} & GitHub and dedicated website \\ \midrule
        \multicolumn{2}{l}{\textbf{Identifier}} & \multicolumn{2}{l}{persistent identifier (journal/ OSF DOI)} & OSHWA unique identifier \\ \midrule
        \multicolumn{2}{l}{\textbf{Metadata}} & OSF and article metadata & article metadata & OSHWA metadata \\ \bottomrule
    \end{tabular}
    \caption{OH use-case examples}
    \label{tab:use-cases}
\end{table}

The Open Science Framework (OSF), as a common choice for disseminating OH design files, was used for the actifield and the thermometer studies. However, while the actifield data contain structured OSF metadata (e.g., license, registration DOI,  tags), the thermometer data deposit lacks structure and has only two folders with code and images. On the other hand, the BioAmp EXG Pill data deposit provides metadata on GitHub (e.g., keywords, license) but without a persistent identifier. Interestingly, some of the actifield data are not shared but available "on reasonable request" ~\cite{KUMBOL2018e00047} even though the study was published in an open-access journal. This presents another complexity in its reuse as the study had closed data, open design files, FOSS, and open documentation.

From the examined use cases, we identify the following challenges of OH reuse: (1) choosing adequate licenses for its design, software, data, and documentation, (2) identifying a dissemination channel (e.g., OSHWA, HardwareX, GitHub, or other), (3) organizing, separating and interlinking resources (e.g., if software and hardware are used for the same purpose researchers tend to choose the same license), and (4) providing detailed metadata and documentation on OH to be reusable, modifiable, and reproducible. Some of these challenges have been previously reported~\cite{bonvoisin2017source, rubow2008open, pearce_return_2016, ackerman2008toward, ohsource} but not fully addressed in practice. In addition, establishing a specialized repository for OH dissemination has not been examined. In the following, we elaborate on the identified challenges.

\subsection{Licensing hardware and software}

The separation of physical and software components of OH has been advised with the instruction to use traditional software licenses for firmware and code loaded in programmable electronic devices~\cite{ayass2012cern, tapr1, cernwiki}. However, schematics with a graphical circuit representation can fall between the analog design and software code categories. They are not solely a connectivity diagram and can serve for circuit simulation and even produce data as, for example, in the Simulation Program with Integrated Circuit Emphasis (SPICE)~\cite{Nagel:M382}. Further, a complexity can emerge when OH is a part of another OH, such as for example, in the non-contact thermometer (licensed under GNU GPL) containing an Arduino UNO unit licensed with Creative Commons (CC BY SA) license~\cite{MNATI2021e00183}. Therefore, licensing OH can be a complex procedure requiring an application-oriented approach rather than defining a common framework~\cite{eurep}. 

Specialized licenses for the OH design component include the CERN Open Hardware License (CERN OHL) with three sharing mechanisms (strongly reciprocal, weakly reciprocal, and permissive), Solderpad, based on the Apache software license, and Tucson Amateur Packet Radio (TAPR) license, adapted from the GNU General Public License. The Solderpad license maintained by the Free and Open Source Silicon Foundation (fossi-foundation.org), has also been recommended as a software license~\cite{soldweb}. Although hardware licenses have existed for about a decade, some of the most successful OH designs (e.g., Arduino) are licensed under Creative Commons (CC) Attribution Share-Alike 2.5 (released in 2005 before the establishment of OH licenses).

\subsection{Dissemination channel selection}

The precondition for OH reuse is that one can access its complete design (i.e., open-source code)~\cite{soldweb}. Public availability of OH is beneficial as sharing provides a firm base for democratic participation in production and reuse~\cite{tsanni2020african}, especially in response to global crises such as the COVID-19 pandemic~\cite{maia2020leveraging, powell_democratizing_2012}. In response to a growing need for medical devices~\cite{maia2020leveraging}, HardwareX journal launched a Special Issue on COVID-19 medical hardware~\cite{speciali} presenting, for example, an open-source solution for non-contact temperature measurement~\cite{MNATI2021e00183}. 

However, there is no universally used repository for open hardware~\cite{bonvoisin2017source}. The three most common dissemination methods are the Open Hardware repository (ohwr.org), git-based repositories (e.g., GitHub, GitLab), and via dedicated journals (i.e., HardwareX, Journal of Open Hardware, Journal of Open Engineering). Certified hardware is typically hosted on GitHub, but some are shared on Google Drive or personal and producer websites and indexed at OSHWA~\cite{oshwac}. HardwareX accepts a variety of repositories for sharing OH and FOSS, including OSF, Mendeley Data, GitHub, and Zenodo. 

Researchers and developers commonly use GitHub to share, store and version their code, data, design files, and documentation. It is based on a well-known distributed version-control system, Git, used for collaborative work in the software developer community. There are numerous benefits of using git-based platforms such as GitHub, GitLab, and BitBucket, but some precautions are warranted. For instance, the current list of officially offered licenses at GitHub does not include any OH licenses. In addition, there is neither a uniformly applicable curation policy nor metadata. FSF constantly evaluates popular repositories and does not encourage using GitHub for several reasons, including the incomplete licensing practice and the use of proprietary software on the platform~\cite{gnugithub}. On the other hand, git-based repositories, including GitHub, are free of charge, convenient to use~\cite{fortunato_case_2021} and effective for version control and collaborative development~\cite{marwick2017computational, lamprecht_towards_2020}. They seamlessly incorporate advanced tools like workflows, software testing, and persistent identifiers~\cite{marwick2017computational, lamprecht_towards_2020}. GitHub has built-in support for repository citations as of August 2021~\cite{githubcit}. Furthermore, git-based platforms can enable FAIR-compliant resource sharing with additional efforts from OH depositors. For instance, the lack of metadata can be bridged with rich documentation in readme files.

Zenodo, Dataverse, and other similar repositories mainly used for sharing publications, data, and software, can also be employed for OH dissemination. Zenodo even supports a deposit type such as Physical Object in the general "Other" category. The main disadvantage of Zenodo is that it does not incorporate software or hardware licenses. Dataverse, on the other hand, specializes in the dissemination of data, meaning that licenses would need to be applied manually. However, both repositories mint a unique identifier DOI, are free to use and have a long-term preservation commitment.

\subsection{Level of documentation detail}

Hardware is generally less documented than software, even though consistent documentation is crucial for complete and accurate OH~\cite{bonvoisin2017source, kera2017science}. Bonvoisin et al. referred to the completeness of OH documentation in terms of the freedoms of FOSS: freedom to study can be exercised by the schematics publication (see Fig.~\ref{fig:arduino} for Arduino MEGA 2560); publishing documents in editable format can support the freedom to modify; freedom to make can be practiced by the publication of bill of materials and assembly instructions, and the selection of appropriate license can grant freedom to distribute. In addition to these principles, it is proposed that documentation of OH should incorporate guidelines for participation, degree of maturity of the shared OH (i.e., design, prototype, full product), and the status of the community (e.g., active or not active)~\cite{bonvoisin2017source}. Although setting an OH documentation standard has been considered critical~\cite{bonvoisin2017source}, it has not materialized, to the best of our knowledge. 

Structured metadata presents an essential aspect of the overall project documentation, especially machine actionable metadata, which is integral to FAIR principles. We address OH metadata in the following section.

\section{FAIR principles for open hardware}

In order to facilitate high-quality research dissemination in the information age, a set of FAIR principles emerged to improve the findability, accessibility, interoperability, and reuse of digital assets. The guidelines emphasized machine-actionability and data management with minimal human intervention due to the ever-increasing complexity and volume of data. The FAIR principles have since been widely recognized and employed in many data repositories and data archiving warehouses. Even though they were primarily intended for data sharing, they found an application in sharing of software code~\cite{lamprecht_towards_2020, peer_challenges_2021}, as code availability was identified as an essential component of scientific reproducibility~\cite{fortunato_case_2021, ince_case_2012}. We argue that OH is the next frontier for FAIR principles.

The first FAIR principle, or \emph{Finable}, mandates that "data and metadata should be easy to find for both humans and computers" ~\cite{community_turing_2019}. OH should be identified with a unique ID. OSHWA certifies open hardware designs and provides a unique identifier, which is not persistent but can be used for this purpose, and commonly used DOIs can be a viable alternative. \emph{Accessibility} mandates access to the resource, potentially with authentication or authorization. It means that OH needs to be retrievable using its digital record via an open, free, and universally used protocol, such as HTTP. The protocol should allow for authentication and authorization when necessary. \emph{Interoperability} means that "data needs to be integrated with other data" and that it "needs to interoperate with applications or workflows for analysis, storage, and processing" ~\cite{community_turing_2019}. Last, the \emph{Reuseable} principle mandates that "metadata and data should be well-described so that they can be replicated and/or combined in different settings" ~\cite{community_turing_2019} for optimal reuse. OH should be described with machine-readable metadata, which is critical for automatically discovering resources on the web. An ongoing Field Ready project (2021-2023) aims to develop and maintain metadata standards for OH.\footnote{https://sloan.org/grant-detail/9626}

Since FAIR principles have been proposed and applied to research data and software, their implementation on OH, which incorporates both, should be attainable. Some aspects of FAIR can be reused from previous work, though there are gaps that need to be addressed for each principle, and an interpretation for OH should be more clearly provided. In Tab.~2 
we show an application of FAIR principles, using "hardware" to denote the digital description of OH (as shown in Fig.~\ref{fig:oh-schema}).

Findability for OH may be implemented with unique identifiers and specific metadata similarly to its implementation for research data and software. Regarding accessibility, we note that infrastructure for OH dissemination remains a challenge as OH files are currently often (disassembled and) shared at multiple places. For interoperability, we emphasize the use of standard knowledge representation and cross-referencing of all required components of OH. Finally, the reusable principle calls for adequate licenses and provenance for all components of OH. Here, we propose a new sub-principle (R2) mandating an explicit dependency tree of OH to other required components, which may include a dependency on other hardware (Tab.~2). 
The sub-principle was modeled on the FAIR for research software framework and further expanded it to a physical realm by adding a reference to available components required for successful OH reuse.

Similar to research data which may not always be released as open data, some specific caveats exist when working with OH. First, source code can be FAIR and shared even if it has proprietary dependencies. For example, the most common source code at the Harvard Dataverse research repository is proprietary~\cite{trisovic2021large}. Similarly, hardware components can comply with the FAIR principles even with proprietary dependencies. Second, some OH medical diagnosis and treatment designs may be limited to research purposes only. Therefore, appropriate regulations and legislation for designs with the intended medical application need to be specified before their dissemination~\cite{maia2020leveraging}.

\clearpage

\begingroup
\footnotesize
\renewcommand{\arraystretch}{1.1} 
\begin{longtable}[c]{>{\raggedright\arraybackslash}p{0.3\linewidth}>{\raggedright\arraybackslash}p{0.3\linewidth}>{\raggedright\arraybackslash}p{0.3\linewidth}}
\caption{FAIR principles for data~\cite{wilkinson_fair_2016}, research software~\cite{katz2021taking}, and open hardware (proposed, modifications are underlined).}
\\
\toprule\toprule
  \textbf{Data (www.go-fair.org)} &
  \textbf{Research software~\cite{katz2021taking}} &
  \textbf{Open Hardware (proposed)} \\ \midrule
\endhead
\multicolumn{3}{c}{\emph{Findable}}\\*
\midrule
F1. (Meta)data are assigned a globally unique and persistent identifier &
F1. \ul{Software is} assigned a globally unique and persistent identifier &
F1. \ul{Hardware is} assigned a globally unique and persistent identifier \ul{through OSHWA or a trusted repository, such that each hardware design and software versions have unique identifier} \\ 
 F2. Data are described with rich metadata (defined by R1 below) &
 F2. \ul{Software is} described with rich metadata &
 F2. \ul{Hardware is} described with rich metadata (defined by R1 below) \\ 
  F3. Metadata clearly and explicitly include the identifier of the data they describe &
  F3. Metadata clearly and explicitly include the identifier of the \ul{software} they describe &
  F3. Metadata clearly and explicitly include the identifier (\ul{DOI or OSHWA}) of the \ul{hardware} they describe \\ 
  F4. (Meta)data are registered or indexed in a searchable resource &
  F4. \ul{Software is} registered or indexed in a searchable resource &
  F4. \ul{Hardware is} registered or indexed in a searchable resource \ul{through OSHWA or a registry} \\ \midrule 
  \multicolumn{3}{c}{\emph{Accessible}}\\*
\midrule\nopagebreak
  A1. (Meta)data are retrievable by their identifier using a standardized communications protocol (the protocol is open, free, and universally implementable, and to allow for an authentication and authorization procedure, where necessary) &
  A1. \ul{Software is} retrievable by \ul{its} identifier using a standardized communications protocol &
  A1. \ul{Hardware is open and} retrievable by \ul{its} identifier using a standardized communications protocol (the protocol is open, free, and universally implementable, and to allow for an authentication and authorization procedure, where necessary). \ul{OH files should be stored cohesively on a repository infrastructure (rather than in multiple disjointed locations), which support long-term hardware stewardship.} \\ 
  A2. Metadata are accessible, even when the data are no longer available. &
  A2. Metadata are accessible, even when the \ul{software is} no longer available &
  A2. Metadata is accessible, even when the \ul{hardware is} no longer available. \\ \midrule
  \multicolumn{3}{c}{\emph{Interoperable}}    \\*
\midrule
  I1. (Meta)data use a formal, accessible, shared, and broadly applicable language for knowledge representation. &
  I1. \ul{Software should read, write or exchange data in a way that meets domain-relevant community standards} &
  I1. \ul{Hardware uses} a formal, accessible, shared, and broadly applicable language for knowledge representation \ul{used in both academia and industry (and enabling their collaboration)}. \\ 
  I2. (Meta)data use vocabularies that follow FAIR principles &
  I2. \ul{Software} includes qualified references to other \ul{objects}. &
  I2. \ul{Hardware uses} vocabularies that follow FAIR principles \\ 
  I3. (Meta)data include qualified references to other (meta)data &
   &
  I3. \ul{Hardware includes cross-references (to own software, data, documentation) and} qualified references to other \ul{objects (e.g., software, data, documentation)}. \\ \midrule\pagebreak
  \multicolumn{3}{c}{\emph{Reusable}}    \\*
\midrule
  R1. (Meta)data are richly described with a plurality of accurate and relevant attributes (with a clear and accessible data usage license, detailed provenance, whilst meeting domain-relevant community standards). &
  R1. \ul{Software is} richly described with a plurality of accurate and relevant attributes &
  
  R1. \ul{Hardware is} richly described with a plurality of accurate and relevant attributes \ul{that reflects its complex structure compliant with the OSHWA definition} (with clear and accessible usage licenses, \ul{to be applied on each of the components and compatible with the dependencies}, detailed provenance \ul{on all components (bill of materials, assembly instructions and other)}, whilst meeting domain-relevant community standards). \\ 
 &
  R2. \ul{Software includes qualified references to other software} &
  R2. \ul{Hardware includes qualified references to other hardware and available components (that would enable reuse)}. \\ \bottomrule
\end{longtable}
\endgroup

\section{Related research}

The CURE-FAIR (CUrating for REproducible FAIR data and code) RDA (Research Data Alliance) working group~\cite{peer_challenges_2021} investigated the current landscape for the adoption of best practices for computational reproducibility by acquiring recommendations and challenges from both literature and the community. The report highlighted the importance of openness and provided some key remarks including FAIR and beyond-FAIR challenges for reproducibility using FOSS, which are also relevant for OH. 



Applying FAIR principles from research data to software turned out to be non-trivial due to the software complexity and unsteadiness~\cite{jimenez2017four, katz2021taking}. Moreover, qualities beyond FAIR such as maintainability of the software, version control, quality control, computational efficacy, and others have been identified as valuable~\cite{lamprecht_towards_2020, peer_challenges_2021}. OH adds to this complexity as, for example, schematics can be seen as both design file and software, and can even produce data. It has been suggested that before defining EU policies on OH and adopting OH widely, appropriate guidelines for OH concerning existing industrial standards should be introduced~\cite{eurep}. Progress has already been made in Germany where the Association of Open Source Ecology Germany (OSEG), in collaboration with the German Institute for Standardization, has carried out a project that aims to develop standardization for OH termed DIN SPEC 3105~\cite{medium1, din1}.

The need for reuse and enhanced findability of existing OH designs has been recognized and addressed in the literature~\cite{EZOJI2021792}. Ezoji et al. conclude that there is a need for good documentation practices that would enable reusability. We believe that applying FAIR principles to OH and adopting good reproducibility practices would enhance the reuse of existing OH designs in research and the industry. 
Besides code availability, an in-detail description of the software, its environment, and hardware requirements should be available to enable reproducible outputs~\cite{ince_case_2012}.

\section{Conclusions}

The paper provides a perspective on leveraging FAIR principles for the dissemination of scientific OH. Considering contemporary science motivations, from research reproducibility to open collaboration, we believe that applying FAIR principles to OH would be a significant step forward, making it documented, finable, accessible, interoperable, and reusable on the web. Thus, it would improve OH curation and its recognition as a complete scientific output. In addition, effective OH dissemination would aid reproducibility on a higher level, beyond the computational part of the study process. FAIR cannot guarantee complete openness of hardware, working functionality, and reproducibility, but it would undoubtedly facilitate it and provide venues for further research across disciplines. Incorporating FAIR principles in the dissemination of scientific OH and adopting best practices such as free OSHWA certification would provide a solid ground for reproducible and reusable research results.

\section{Acknowledgments}

AT is funded by Alfred P. Sloan Foundation (grant number P-2020-13988). NM was partly supported by Grant No. TR33020 funded by the Ministry of Education, Science and Technological Development, Republic of Serbia.

The authors would like to thank CURE-FAIR members for their valuable contribution to the RDA community.

\bibliographystyle{IEEEtran}

\begingroup
\raggedright
\bibliography{PSSOH-rad}

\begin{thebibliography}{10}
\providecommand{\url}[1]{#1}
\csname url@samestyle\endcsname
\providecommand{\newblock}{\relax}
\providecommand{\bibinfo}[2]{#2}
\providecommand{\BIBentrySTDinterwordspacing}{\spaceskip=0pt\relax}
\providecommand{\BIBentryALTinterwordstretchfactor}{4}
\providecommand{\BIBentryALTinterwordspacing}{\spaceskip=\fontdimen2\font plus
\BIBentryALTinterwordstretchfactor\fontdimen3\font minus
  \fontdimen4\font\relax}
\providecommand{\BIBforeignlanguage}[2]{{%
\expandafter\ifx\csname l@#1\endcsname\relax
\typeout{** WARNING: IEEEtran.bst: No hyphenation pattern has been}%
\typeout{** loaded for the language `#1'. Using the pattern for}%
\typeout{** the default language instead.}%
\else
\language=\csname l@#1\endcsname
\fi
#2}}
\providecommand{\BIBdecl}{\relax}
\BIBdecl

\bibitem{dryden_upon_2017}
M.~D.~M. Dryden, R.~Fobel, C.~Fobel, and A.~R. Wheeler, ``Upon the {Shoulders}
  of {Giants}: {Open}-{Source} {Hardware} and {Software} in {Analytical}
  {Chemistry},'' \emph{Analytical Chemistry}, vol.~89, no.~8, pp. 4330--4338,
  Apr. 2017, publisher: American Chemical Society.

\bibitem{hocquet_epistemic_2021}
\BIBentryALTinterwordspacing
A.~Hocquet and F.~Wieber, ``\BIBforeignlanguage{en}{Epistemic issues in
  computational reproducibility: software as the elephant in the room},''
  \emph{\BIBforeignlanguage{en}{European Journal for Philosophy of Science}},
  vol.~11, no.~2, p.~38, Jun. 2021. [Online]. Available:
  \url{https://link.springer.com/10.1007/s13194-021-00362-9}
\BIBentrySTDinterwordspacing

\bibitem{wheeler2011free}
D.~A. Wheeler, ``Why free-libre/open source software (floss)? look at the
  numbers!'' 2011.

\bibitem{fortunato_case_2021}
L.~Fortunato and M.~Galassi, ``\BIBforeignlanguage{en}{The case for free and
  open source software in research and scholarship},''
  \emph{\BIBforeignlanguage{en}{Philosophical Transactions of the Royal Society
  A: Mathematical, Physical and Engineering Sciences}}, vol. 379, no. 2197, pp.
  rsta.2020.0079, 20\,200\,079, May 2021.

\bibitem{marwick2017computational}
B.~Marwick, ``Computational reproducibility in archaeological research: Basic
  principles and a case study of their implementation,'' \emph{Journal of
  Archaeological Method and Theory}, vol.~24, no.~2, pp. 424--450, 2017.

\bibitem{martinez2018reproducibility}
C.~Martinez, J.~Hollister, B.~Marwick, E.~Sz{\"o}cs, S.~Zeitlin, B.~Kinoshita,
  and B.~Meinke, ``Reproducibility in science: A guide to enhancing
  reproducibility in scientific results and writing,'' \emph{URL
  http://ropensci. github. io/reproducibility-guide}, 2018.

\bibitem{community_turing_2019}
\BIBentryALTinterwordspacing
T.~T.~W. Community, B.~Arnold, L.~Bowler, S.~Gibson, P.~Herterich, R.~Higman,
  A.~Krystalli, A.~Morley, M.~O'Reilly, and K.~Whitaker, ``The {Turing} {Way}:
  {A} {Handbook} for {Reproducible} {Data} {Science},'' Mar. 2019. [Online].
  Available: \url{https://zenodo.org/record/3233986}
\BIBentrySTDinterwordspacing

\bibitem{gosh}
{Gathering for Open Science Hardware (GOSH)}, ``{GOSH Manifesto},''
  \url{https://openhardware.science/gosh-manifesto/#GOSH_makes_science_better},
  accessed: 2021-09-08.

\bibitem{freesw}
``{What is Free Software?}'' \url{https://www.gnu.org/philosophy/free-sw.html},
  accessed: 2021-09-08.

\bibitem{stallman2002free}
R.~Stallman, \emph{Free software, free society: Selected essays of Richard M.
  Stallman}.\hskip 1em plus 0.5em minus 0.4em\relax Lulu. com, 2002.

\bibitem{tapr1}
``{The TAPR Open Hardware License},''
  \url{https://tapr.org/the-tapr-open-hardware-license/}, accessed: 2021-09-01.

\bibitem{ohsource}
``{What is open hardware?}''
  \url{https://opensource.com/resources/what-open-hardware}, accessed:
  2021-09-08.

\bibitem{bonvoisin2017source}
J.~Bonvoisin, R.~Mies, J.-F. Boujut, and R.~Stark, ``What is the “source”
  of open source hardware?'' 2017.

\bibitem{pearce_return_2016}
J.~M. Pearce, ``\BIBforeignlanguage{en}{Return on investment for open source
  scientific hardware development},'' \emph{\BIBforeignlanguage{en}{Science and
  Public Policy}}, vol.~43, no.~2, pp. 192--195, Apr. 2016.

\bibitem{rubow2008open}
E.~Rubow, ``Open source hardware,'' \emph{T ech. rep}, pp. 1--5, 2008.

\bibitem{ackerman2008toward}
J.~R. Ackerman, ``Toward open source hardware,'' \emph{U. Dayton L. Rev.},
  vol.~34, p. 183, 2008.

\bibitem{HEIKKINEN2020119986}
I.~Heikkinen, H.~Savin, J.~Partanen, J.~Sepp{\"a}l{\"a}, and J.~Pearce,
  ``Towards national policy for open source hardware research: The case of
  finland,'' \emph{Technological Forecasting and Social Change}, vol. 155, p.
  119986, 2020.

\bibitem{stodden_enabling_2018}
V.~Stodden, M.~S. Krafczyk, and A.~Bhaskar, ``\BIBforeignlanguage{en}{Enabling
  the {Verification} of {Computational} {Results}: {An} {Empirical}
  {Evaluation} of {Computational} {Reproducibility}},'' in
  \emph{\BIBforeignlanguage{en}{Proceedings of the {First} {International}
  {Workshop} on {Practical} {Reproducible} {Evaluation} of {Computer}
  {Systems}}}.\hskip 1em plus 0.5em minus 0.4em\relax Tempe AZ USA: ACM, Jun.
  2018, pp. 1--5.

\bibitem{engineering2019reproducibility}
{Engineering, Medicine and on Behavioral, Board and National Academies of
  Sciences, Engineering, and Medicine and others}, ``{Reproducibility and
  Replicability in Science},'' 2019.

\bibitem{peer_challenges_2021}
\BIBentryALTinterwordspacing
L.~Peer, F.~Arguillas, T.~Honeyman, N.~Miljković, K.~P.-V. Gehlen, and C.-F.
  W.~S. 3, ``\BIBforeignlanguage{en}{Challenges of {Curating} for
  {Reproducible} and {FAIR} {Research} {Output}},'' 2021, publisher: Research
  Data Alliance Version Number: 2.1. [Online]. Available:
  \url{https://zenodo.org/record/5094155#.YO0a8OgzaUk}
\BIBentrySTDinterwordspacing

\bibitem{wilkinson_fair_2016}
\BIBentryALTinterwordspacing
M.~D. Wilkinson, M.~Dumontier, I.~J. Aalbersberg, G.~Appleton, M.~Axton,
  A.~Baak, N.~Blomberg, J.-W. Boiten, L.~B. da~Silva~Santos, P.~E. Bourne,
  J.~Bouwman, A.~J. Brookes, T.~Clark, M.~Crosas, I.~Dillo, O.~Dumon,
  S.~Edmunds, C.~T. Evelo, R.~Finkers, A.~Gonzalez-Beltran, A.~J. Gray,
  P.~Groth, C.~Goble, J.~S. Grethe, J.~Heringa, P.~A. ’t Hoen, R.~Hooft,
  T.~Kuhn, R.~Kok, J.~Kok, S.~J. Lusher, M.~E. Martone, A.~Mons, A.~L. Packer,
  B.~Persson, P.~Rocca-Serra, M.~Roos, R.~van Schaik, S.-A. Sansone,
  E.~Schultes, T.~Sengstag, T.~Slater, G.~Strawn, M.~A. Swertz, M.~Thompson,
  J.~van~der Lei, E.~van Mulligen, J.~Velterop, A.~Waagmeester, P.~Wittenburg,
  K.~Wolstencroft, J.~Zhao, and B.~Mons, ``\BIBforeignlanguage{en}{The {FAIR}
  {Guiding} {Principles} for scientific data management and stewardship},''
  \emph{\BIBforeignlanguage{en}{Scientific Data}}, vol.~3, no.~1, p. 160018,
  Dec. 2016. [Online]. Available:
  \url{http://www.nature.com/articles/sdata201618}
\BIBentrySTDinterwordspacing

\bibitem{pejovic_predrag_2021_4748368}
P.~Pejovi\'{c}, N.~Miljkovi\'{c}, M.~Cvetanovi\'{c}, and
  M.~\v{S}evku\v{s}i\'{c}, ``Licence slobodnog softvera i otvorenog hardvera -
  kratko uputstvo za nestrpljive.''\hskip 1em plus 0.5em minus 0.4em\relax
  Belgrade, Serbia: {University of Belgrade - School of Electrical Engineering
  and Academic Mind}, May 2021, {Title in English: ``Free Software and Open
  Hardware Licenses: A Short Guide for People in a Hurry''. This record
  contains both paper and presentation.}

\bibitem{eurep}
P.~Kauttu, L.~F.~R. Murillo, L.~Pujol~Priego, J.~Wareham, and A.~Katz, ``{Open
  hardware licences: Parallels and contrasts},'' European Commission, Tech.
  Rep., 2019.

\bibitem{frnational}
``{Second National Plan for Open Science},''
  \url{https://www.ouvrirlascience.fr/second-national-plan-for-open-science/},
  accessed: 2021-09-08.

\bibitem{cernoh}
``{CERN launches Open Hardware initiative},''
  \url{https://phys.org/news/2011-07-cern-hardware.html}, accessed: 2021-09-08.

\bibitem{mega25}
``{Arduino Mega 2560},''
  \url{https://content.arduino.cc/assets/MEGA2560_Rev3e_sch.pdf}, accessed:
  2021-09-08.

\bibitem{requiredbrain}
\BIBentryALTinterwordspacing
``{Arduino: History and Success of Open Source Electronic Micro-controllers},''
  2021, accessed 30 August 2021. [Online]. Available:
  \url{https://www.requiredbrain.com/arduino-history-and-success-of-open-source-electronic-micro-controllers}
\BIBentrySTDinterwordspacing

\bibitem{arduino1}
``{Arduino Store:} arduino mega 2560 rev3,''
  \url{https://store.arduino.cc/products/arduino-mega-2560-rev3}, accessed:
  2021-09-01.

\bibitem{pearce2012building}
J.~M. Pearce, ``Building research equipment with free, open-source hardware,''
  \emph{Science}, vol. 337, no. 6100, pp. 1303--1304, 2012.

\bibitem{KUMBOL2018e00047}
``Actifield, an automated open source actimeter for rodents,''
  \emph{HardwareX}, vol.~4, p. e00047, 2018.

\bibitem{MNATI2021e00183}
``An open-source non-contact thermometer using low-cost electronic
  components,'' \emph{HardwareX}, vol.~9, p. e00183, 2021.

\bibitem{bioamp}
``Bioamp exg pill,'' \url{https://github.com/upsidedownlabs/BioAmp-EXG-Pill},
  accessed: 2021-09-08.

\bibitem{todd_duncombe_2021_5078227}
T.~Duncombe, ``Hardwarex manuscript templates,''
  \url{https://doi.org/10.5281/zenodo.5078227}, Jul. 2021.

\bibitem{ayass2012cern}
M.~Ayass and J.~Serrano, ``The cern open hardware license,'' \emph{IFOSS L.
  Rev.}, vol.~4, p.~71, 2012.

\bibitem{cernwiki}
``{CERN Open Hardware Licence Wiki},''
  \url{https://ohwr.org/project/cernohl/wikis/faq}, accessed: 2021-09-08.

\bibitem{Nagel:M382}
L.~W. Nagel and D.~Pederson, ``Spice (simulation program with integrated
  circuit emphasis),'' EECS Department, University of California, Berkeley,
  Tech. Rep. UCB/ERL M382, Apr 1973.

\bibitem{soldweb}
``{Solderpad Website},''
  \url{https://github.com/fossi-foundation/solderpad-website/blob/master/index.md},
  accessed: 2021-09-08.

\bibitem{tsanni2020african}
A.~Tsanni, ``African scientists leverage open hardware,'' \emph{Nature}, vol.
  582, no. 7810, pp. 138--139, 2020.

\bibitem{maia2020leveraging}
A.~Maia~Chagas, J.~C. Molloy, L.~L. Prieto-Godino, and T.~Baden, ``Leveraging
  open hardware to alleviate the burden of covid-19 on global health systems,''
  \emph{PLoS biology}, vol.~18, no.~4, p. e3000730, 2020.

\bibitem{powell_democratizing_2012}
A.~Powell, ``\BIBforeignlanguage{en}{Democratizing production through open
  source knowledge: from open software to open hardware},''
  \emph{\BIBforeignlanguage{en}{Media, Culture \& Society}}, vol.~34, no.~6,
  pp. 691--708, Sep. 2012.

\bibitem{speciali}
``Special issue : Covid-19 medical hardware,''
  \url{https://www.sciencedirect.com/journal/hardwarex/special-issue/1037TX0Z2W8},
  2021, accessed: 2021-09-08.

\bibitem{oshwac}
``Certified open-source hardware projects,''
  \url{https://certification.oshwa.org/list.html}, accessed: 2021-09-08.

\bibitem{gnugithub}
``{GNU Ethical Repository Criteria Evaluations},''
  \url{https://www.gnu.org/software/repo-criteria-evaluation.html#GitHub},
  accessed: 2021-09-08.

\bibitem{lamprecht_towards_2020}
A.-L. Lamprecht, L.~Garcia, M.~Kuzak, C.~Martinez, R.~Arcila, E.~Martin
  Del~Pico, V.~Dominguez Del~Angel, S.~van~de Sandt, J.~Ison, P.~A. Martinez,
  P.~McQuilton, A.~Valencia, J.~Harrow, F.~Psomopoulos, J.~L. Gelpi,
  N.~Chue~Hong, C.~Goble, and S.~Capella-Gutierrez, ``Towards {FAIR} principles
  for research software,'' \emph{Data Science}, vol.~3, no.~1, pp. 37--59, Jun.
  2020.

\bibitem{githubcit}
A.~Smith, ``{Enhanced support for citations on GitHub},''
  \url{https://github.blog/2021-08-19-enhanced-support-citations-github/},
  accessed: 2021-09-08.

\bibitem{kera2017science}
D.~Kera, ``Science artisans and open science hardware,'' \emph{Bulletin of
  Science, Technology \& Society}, vol.~37, no.~2, pp. 97--111, 2017.

\bibitem{ince_case_2012}
D.~C. Ince, L.~Hatton, and J.~Graham-Cumming, ``\BIBforeignlanguage{en}{The
  case for open computer programs},'' \emph{\BIBforeignlanguage{en}{Nature}},
  vol. 482, no. 7386, pp. 485--488, Feb. 2012.

\bibitem{trisovic2021large}
A.~Trisovic, M.~K. Lau, T.~Pasquier, and M.~Crosas, ``A large-scale study on
  research code quality and execution,'' \emph{arXiv preprint
  arXiv:2103.12793}, 2021.

\bibitem{katz2021taking}
D.~S. Katz, M.~Gruenpeter, and T.~Honeyman, ``Taking a fresh look at fair for
  research software,'' \emph{Patterns}, vol.~2, no.~3, p. 100222, 2021.

\bibitem{jimenez2017four}
R.~C. Jim{\'e}nez, M.~Kuzak, M.~Alhamdoosh, M.~Barker, B.~Batut, M.~Borg,
  S.~Capella-Gutierrez, N.~C. Hong, M.~Cook, M.~Corpas \emph{et~al.}, ``Four
  simple recommendations to encourage best practices in research software,''
  \emph{F1000Research}, vol.~6, 2017.

\bibitem{medium1}
M.~Kollotzek, ``Adapting open source methods to products,''
  \url{https://medium.com/open-culture-journal/adapting-open-source-methods-to-products-af08bd4b4643},
  2019, accessed: 2021-09-01.

\bibitem{din1}
``{DIN SPEC 3105} open source hardware,''
  \url{https://www.din.de/en/wdc-beuth:din21:305669958?destinationLanguage=&sourceLanguage=},
  accessed: 2021-09-01.

\bibitem{EZOJI2021792}
A.~Ezoji, J.~Boujut, and R.~Pinqui{\'e}, ``Requirements for design reuse in
  open-source hardware: a state of the art,'' \emph{Procedia CIRP}, vol. 100,
  pp. 792--797, 2021, 31st CIRP Design Conference 2021 (CIRP Design 2021).

\end{thebibliography}
\endgroup
\end{document}